\documentclass[twocolumn]{aastex62}

\usepackage{apjfonts}
\usepackage{amsmath}

\newcommand{\Hi}{\hbox{H{\sc i}}}

\begin{document}

\title{The inflow and outflow rate evolution of local Milky Way-mass star-forming galaxies since $z=1.3$}
\author{Zhizheng Pan}
\email{panzz@pmo.ac.cn, xzzheng@pmo.ac.cn, xkong@ustc.edu.cn}
\affiliation{Purple Mountain Observatory, Chinese Academy of Sciences, 8 Yuan Hua Road, Nanjing, Jiangsu 210008, China}
\affiliation{School of Astronomy and Space Sciences, University of Science and Technology of China, Hefei, 230026, China
}

\author{Yingjie Peng}
\affiliation{Kavli Institute for Astronomy and Astrophysics, Peking University, Yi He Yuan Lu 5, Hai Dian District, Beijing 100871, China}

\author{Xianzhong Zheng}
\affiliation{Purple Mountain Observatory, Chinese Academy of Sciences, 8 Yuan Hua Road, Nanjing, Jiangsu 210008, China}
\affiliation{School of Astronomy and Space Sciences, University of Science and Technology of China, Hefei, 230026, China
}

\author{Jing Wang}
\affiliation{Kavli Institute for Astronomy and Astrophysics, Peking University, Yi He Yuan Lu 5, Hai Dian District, Beijing 100871, China}

\author{Xu Kong}
\affiliation{School of Astronomy and Space Sciences, University of Science and Technology of China, Hefei, 230026, China
}
\affiliation{CAS Key Laboratory for Research in Galaxies and Cosmology, Department of Astronomy, \\
University of Science and Technology of China, Hefei, Anhui 230026, China}

\begin{abstract}
We study the gas inflow rate ($\zeta_{\rm inflow}$) and outflow rate ($\zeta_{\rm outflow}$) evolution of local Milky Way-mass star-forming galaxies (SFGs) since $z=1.3$. The stellar mass growth history of Milky Way-mass progenitor SFGs is inferred from the evolution of the star formation rate (SFR)$-$stellar mass ($M_{\ast}$) relation, and the gas mass ($M_{\rm gas}$) is derived using the recently established gas scaling relations. With the $M_{\ast}+M_{\rm gas}$ growth curve, the net inflow rate $\kappa$ is quantified at each cosmic epoch. At $z\sim 1.3$, $\kappa$ is comparable with the SFR, whereas it rapidly decreases to $\sim 0.15\times$SFR at $z=0$.  We then constrain the average outflow rate $\zeta_{\rm outflow}$ of progenitor galaxies by modeling the evolution of their gas-phase metallicity. The best-fit $\zeta_{\rm outflow}$ is found to be $(0.5-0.8)\times$SFR. Combining $\kappa$ and $\zeta_{\rm outflow}$, we finally investigate the evolution of $\zeta_{\rm inflow}$ since $z=1.3$. We find that $\zeta_{\rm inflow}$ rapidly decreases by $\sim$80\% from $z=1.3$ to $z=0.5$. At $z<0.5$, $\zeta_{\rm inflow}$ continuously decreases but with a much lower decreasing rate. Implications of these findings on galaxy evolution are discussed.
\end{abstract}

\section{Introduction} \label{sec:intro}
In the current galaxy formation paradigm, gas flows into and out of galaxies are key ingredients for driving galaxy evolution \citep{Bouche 2010,Dave 2011, Dave 2012,Lilly 2013, Peng 2014}. Observational studies suggest that gas inflows are required for SFGs, as their gas depletion time scale is significantly shorter than that required to build up their stellar mass in both the low-redshift and high-redshift universe \citep{Larson 1980,Genzel 2015,Tacconi 2018}. As an important feedback mechanism, gas outflows driven by star formation or active galactic nucleus (AGNs) can blow the metal-enriched gas out of a galaxy, regulating its chemical enrichment and star formation \citep{Peeples 2011,Hopkins 2012,Cicone 2014,Geach 2014}.

Theoretical works have predicted that gas inflows are achieved in two different modes, which are termed as the ``cold mode" and the ``hot mode" accretion \citep[e.g.,][]{Keres 2005,Dekel 2006}. In low-mass halos and high redshift universe, gas is acquired primarily through the cold mode accretion, by which cold gas flows can directly feed galaxies through cosmic filaments \citep{Keres 2005,Dekel 2009a, Dekel 2009b,van de Voort 2011}. When a galaxy's dark matter halo grows massive enough to support a stable shock, the infalling gas is first shock-heated to near the viral temperature ($T\sim 10^{6}K$), then radiatively cools and settles into galaxies in a quasi-spherical manner. The transition of these two accretion modes is expected to occur near the critical halo mass, $M_{\rm c}\sim 10^{12}M_{\sun}$ \citep{Dekel 2006}. To justify this, it is important to investigate the behavior of gas accretion when a galaxy evolves across $M_{\rm c}$. Simulations suggest that the gas accretion behavior indeed changes near $M_{\rm c}$ \citep{Stewart 2011}, but observational confirmation of this is still lacking.

Observationally, gas flow signatures have been unambiguously detected in the high-quality spectra of SFGs \citep[e.g.,][]{Heckman 1990,Sato 2009,Weiner 2009,Genzel 2014a,Rubin 2014,Cicone 2016}. Nevertheless, the detailed properties of gas flows are still difficult to quantify directly. This is because gas flows can occur in multi-phase, and the global gas flow rates depend on the 3D motions and densities of the gas. Indirect methods are thus useful in studying gas flows. For example, early attempts have tried to set constraints on gas flows by modeling the chemical evolution of SFGs to match the observed  mass-metallicity relation \citep{Finlator 2008,Spitoni 2010,Lilly 2013,Yabe 2015,Spitoni 2017}.

The assembly history of Milky Way-mass ($M_{\rm MW}\sim 5\times 10^{10}M_{\sun}$, see \citealt{McMillan 2017}) galaxies has recently attracted much attentions, since galaxies near $M_{\rm MW}$ appear quite typical and dominate the stellar mass budget in the local Universe \citep{vanDokkum 2013}. Several works have tried to trace the evolution of star formation and morphology of $M_{\rm MW}$ progenitor galaxies back to $z=1-2$ \citep{Patel 2013,vanDokkum 2013,Papovich 2015}. In this paper, we aim to study the gas inflow and outflow history of local $M_{\rm MW}$ SFGs using an indirect approach. In Section 2, we first use the technique developed by \citet{Leitner 2011} to select $M_{\rm MW}$ progenitor SFGs up to $z=1.3$.  In Section 3, we infer the molecular gas mass ($M_{\rm H2}$) of progenitor galaxies using the scaling relation recently established by \citet{Tacconi 2018}, and the atomic gas mass ($M_{\rm HI}$) is inferred using the $M_{\rm HI}-M_{\ast}$ relation established at $z=0$. In Section 4, we quantify the net inflow rate evolution of progenitor galaxies with the $M_{\ast}+M_{\rm gas}$ growth curve. In Section 5, we use an analytical chemical evolution model to set constraints on the outflow rate of progenitor galaxies. With the derived net inflow rate and outflow rate, we can investigate the gas inflow history of the local $M_{\rm MW}$ SFGs. In Section 6, we discuss the implication of our results. Finally, we summarize our findings in Section 7. Throughout this paper, we adopt a concordance $\Lambda$CDM cosmology with $\Omega_{\rm m}=0.3$, $\Omega_{\rm \Lambda}=0.7$, $H_{\rm 0}=70$ $\rm km~s^{-1}$ Mpc$^{-1}$ and a \citet{Chabrier 2003} initial mass function (IMF). All reported gas masses in this work include a correction of 1.36 to account for helium.

\begin{figure}
\centering
\includegraphics[width=80mm,angle=0]{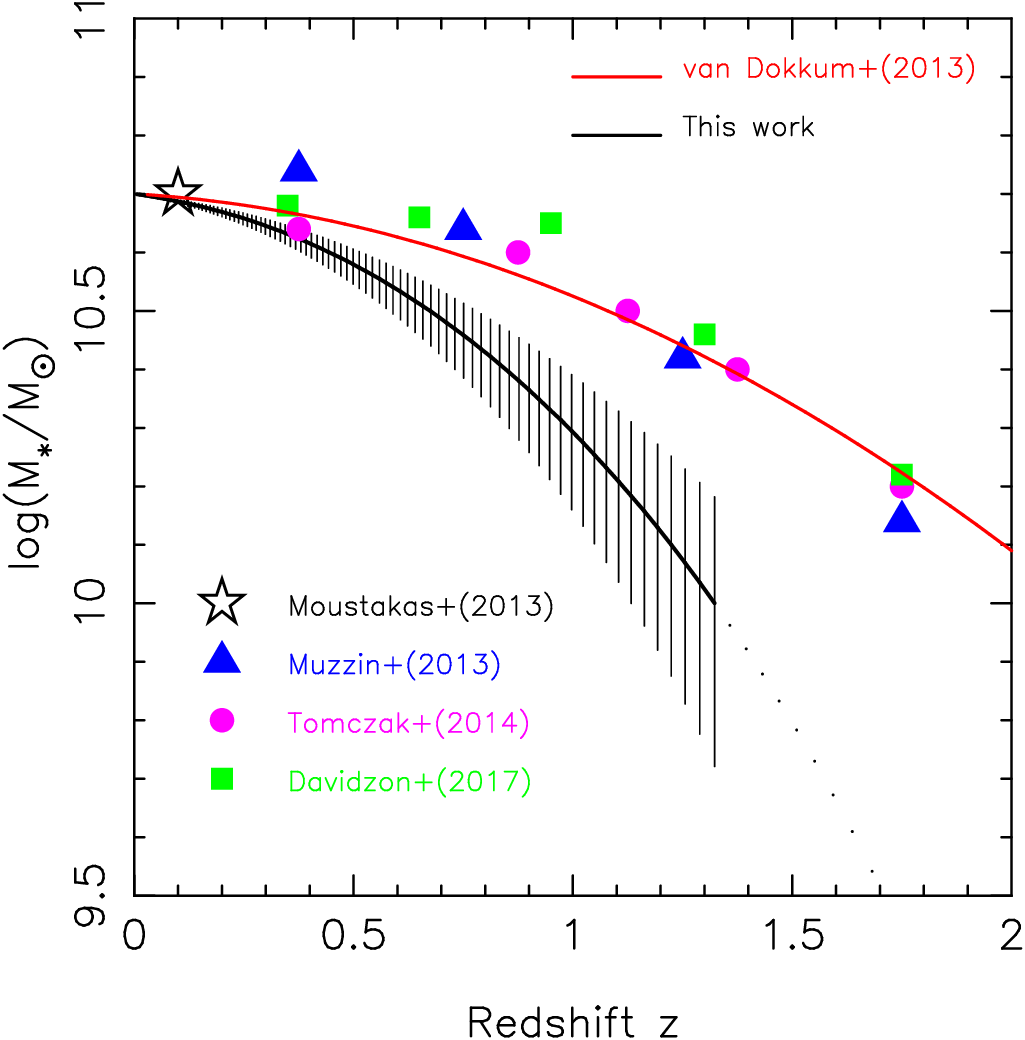}
\caption{The black solid curve shows the stellar mass growth history of a SFG that with log$(M_{\ast}/M_{\sun})=10.7$ at $z=0$.  Considering the different main sequence parameterizations of different works, we include a $\pm0.1$ dex variation in the main sequence parameterized in equation (1). The resulting uncertainty in the stellar mass growth history is shown in the hatched region. The inferred stellar mass growth history becomes increasingly uncertain towards high redshift. We also show the mass growth history given by \citet{vanDokkum 2013} (red curve), who select the Milky Way-mass progenitor galaxies with a constant cumulative comoving number density of $\rho_c=1.1\times10^{-3}\rm Mpc^{-3}$. The large symbols indicate the stellar mass at which the cumulative comoving density reaches $\rho_c=1.1\times10^{-3}\rm Mpc^{-3}$, which are drawn from some recently published stellar mass functions.}\label{fig1}
\end{figure}

\section{Stellar mass growth history of progenitor galaxies}
We use the method developed by \citet{Leitner 2011}, namely the Main Sequence Integration (MSI) approach, to select progenitors of SFGs that with final stellar mass of log($M_{\ast}/M_{\sun})=10.7$. The philosophy of this method is simple: if SFGs assemble most of their stellar mass from in situ star formation, then for a given redshift interval, the new stellar mass added to the existing mass is computable based on the location of the galaxy on the SFR$-M_{\ast}$ plane and mass loss from stellar evolution modeling. For local $M_{\rm MW}$ SFGs, this method should be valid as galaxies with stellar mass near or below $M_{\rm MW}$ assemble their mass mainly from in situ star formation, not from mergers \citep{Qu 2017,Behroozi 2018}. From the observational perspective, the assumption that local $M_{\rm MW}$ SFGs are always star-forming in the past is supported by the stellar population constituents of the Milky Way disk \citep{Haywood 2016}. Details of the MSI approach can be found in \citet{Leitner 2011}.

The SFR$-M_{\ast}$ relation we used is from the work of \citet{speagle 2014}, in which the evolution of the SFR$-M_{\ast}$ relation at $z=[0,6]$ is systematically investigated based on the compiled data from 25 studies. At each cosmic epoch, the SFR$-M_{\ast}$ relation can be characterized by:
\begin{equation}
{\rm log~}{\rm SFR}(M_{\ast},t)=(0.84-0.026\times t){\rm log}M_{\ast}-(6.51-0.11\times t),
\end{equation}
where $t$ is the age of the universe in Gyr. Note that at a given $t$, the SFR$-M_{\ast}$ relation is parameterized by a single power-law. This may be problematic since the SFR$-M_{\ast}$ relation appears having different power-law indices in the low- and high- mass regimes, as reported in some recent studies \citep{Whitaker 2014, Lee 2015, Schreiber 2015, Tomczak 2016}. To investigate whether equation (1) is a good description of the star formation main sequence, we have compared the SFR$-M_{\ast}$ relation of \citet{speagle 2014} with those of \cite{Whitaker 2014} and \citet{Tomczak 2016}. At the same cosmic epoch, we find that these works report a remarkable consistent SFR$-M_{\ast}$ relation at log$(M_{\ast}/M_{\sun})=[10.0,11.0]$, with a typical discrepancy of $\Delta {\rm log~SFR}<0.05$ dex at fixed $M_{\ast}$. Since in this work we only trace progenitor galaxies back to $z=1.3$ where they have a stellar mass of log$(M_{\ast}/M_{\sun})=10.0$, the stellar mass growth history inferred from equation (1) should be robust.

In Figure \ref{fig1}, we show the mass growth history of progenitor galaxies. To account for the uncertainty of the evolution of the main sequence, we arbitrarily allow a $\pm0.1$ dex variation in the star formation rate ($\Delta {\rm log(SFR)=0.1}$). The resulting uncertainty in the stellar mass growth history ($\Delta {\rm log}M_{\ast}$) is shown in the hatched region. When increasing $\Delta \rm{log(SFR)=0.1}$ to $\Delta \rm{log(SFR)=0.3}$, $\Delta {\rm log}M_{\ast}$ increases by a factor of $\rm 5$ and $\sim 1.5$ at $z\sim 1.3$ and $z\sim 0.5$, respectively.  For comparison, we also show the mass growth history of Milky Way-mass galaxies presented by \citet{vanDokkum 2013}, who select progenitor galaxies with a constant cumulative comoving number density of $\rho_c=1.1\times10^{-3}~\rm Mpc^{-3}$.  It is clear from Figure \ref{fig1} that the number density selection method is always biased to select more massive galaxies. This is because \citet{vanDokkum 2013} also select the progenitors of quiescent Milky Way-mass galaxies. With a same final mass, it is natural that the quiescent ones will always assemble much earlier than the star-forming ones. It is worthy to note that the inferred mass growth history of progenitor galaxies becomes increasingly uncertain towards high redshifts. Given this, in the following we only focus on the evolution of progenitor galaxies at $z<1.3$.

\section{The determination of cold gas mass $M_{\rm gas}$}
The cold gas component of a galaxy consists of molecular and atomic hydrogen ($\rm H_{\rm 2}$ and \Hi). Thanks to the increasing galaxy sample collected by recent molecular gas surveys \cite[e.g.,][]{Daddi 2010, Tacconi 2010,Tacconi 2013,Saintonge 2011,Saintonge 2017,Combes 2011}, scaling relations between molecular gas mass ($M_{\rm H2}$), redshifts and star formation rates for SFGs are now established. The seminal work of \citet{Genzel 2015} compiled data from a number of molecular gas surveys at $z=[0, 3]$ to establish scaling relations between $\tau_{\rm H2}$ (the $\rm H_{\rm 2}$ depletion time scale, defined as $\tau_{\rm H2}=M_{\rm H2}/$SFR), $M_{\ast}$, SFR and redshift $z$, enabling the determination of $M_{\rm H2}$ for SFGs to an accuracy of $\pm$0.2 dex. Recently, \citet{Tacconi 2018} updated and improved the scaling relations of \citet{Genzel 2015} using a larger sample spanning $z=[0, 4]$. With the new scaling relations, it is possible to determine $\tau_{\rm H2}$ (or $M_{\rm H2}$) to an accuracy of $\pm 0.1$ dex or better for sample averages. For SFGs that lie on the ridge line of the SFR$-M_{\ast}$ relation of \citet{speagle 2014}, the dependence of $\tau_{\rm H2}$ on redshift can be characterized by:
\begin{equation}
{\rm log}(\tau_{\rm H2})=0.09-0.62{\rm log}(1+z)~~~~~~~\rm Gyr,
\end{equation}
where $z$ is redshift. As shown in \citet{Tacconi 2018}, $\tau_{\rm H2}$ shows no clear dependence on $M_{\ast}$, at least at  log$(M_{\ast}/M_{\sun})>10.0$. Therefore, we can infer the $M_{\rm H2}$ of progenitor galaxies at any cosmic epoch with:
\begin{equation}
{\rm log}(M_{\rm H_{2}}/M_{\sun})=\rm {log(SFR)}+{\rm log}(\tau_{\rm H2}).
\end{equation}

\begin{figure}
\centering
\includegraphics[width=80mm,angle=0]{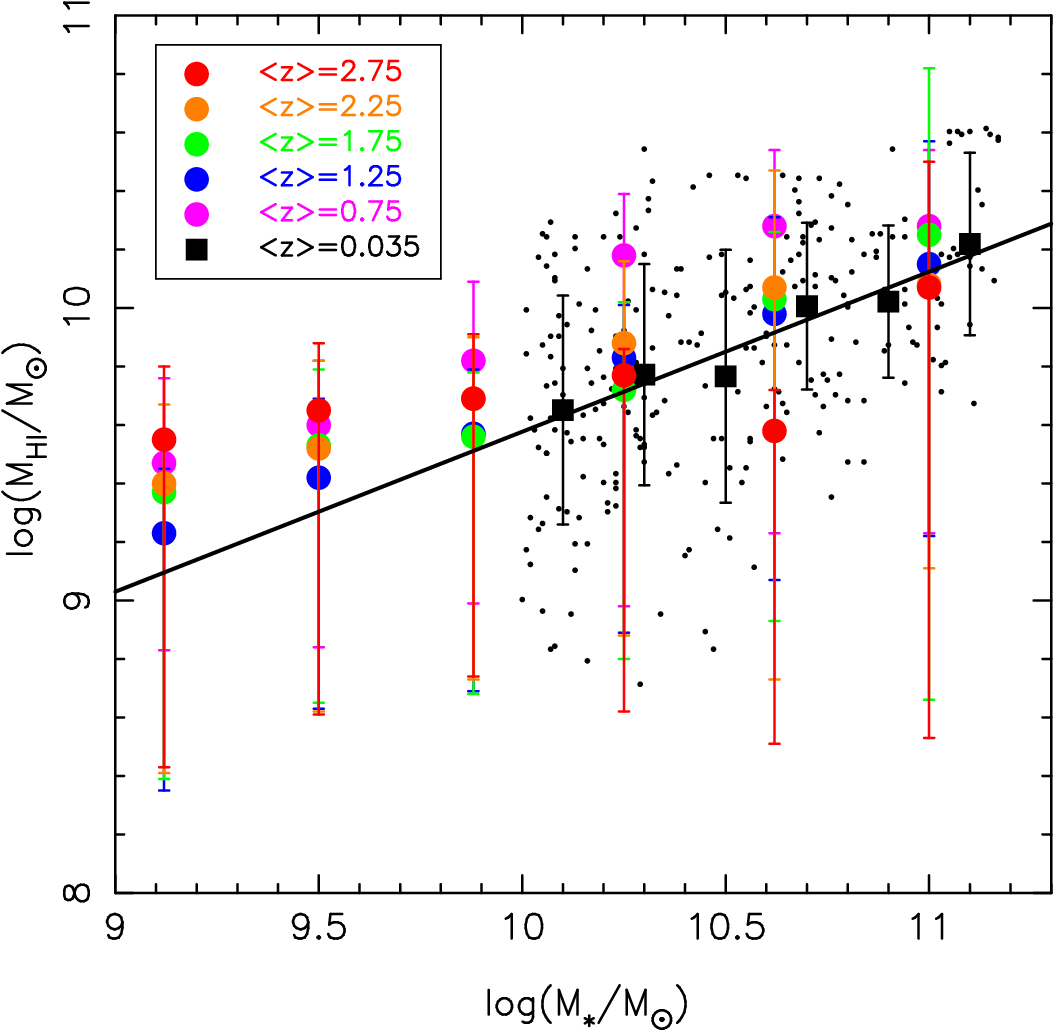}
\caption{The log$(M_{\rm HI})-{\rm log}(M_{\ast})$ relation of SFGs. Small black symbols are SFGs with direct \Hi~measurements from the GASS survey, with definite \Hi~detection and NUV$-r<4.0$. Large color symbols are from \citet{Popping 2015}. The running median of the GASS galaxies are shown in black squares. The black solid line is the linear fit of the GASS sample. Error bars indicate the $1\sigma$ scatter.}\label{fig2}
\end{figure}

\begin{figure*}
\centering
\includegraphics[width=160mm,angle=0]{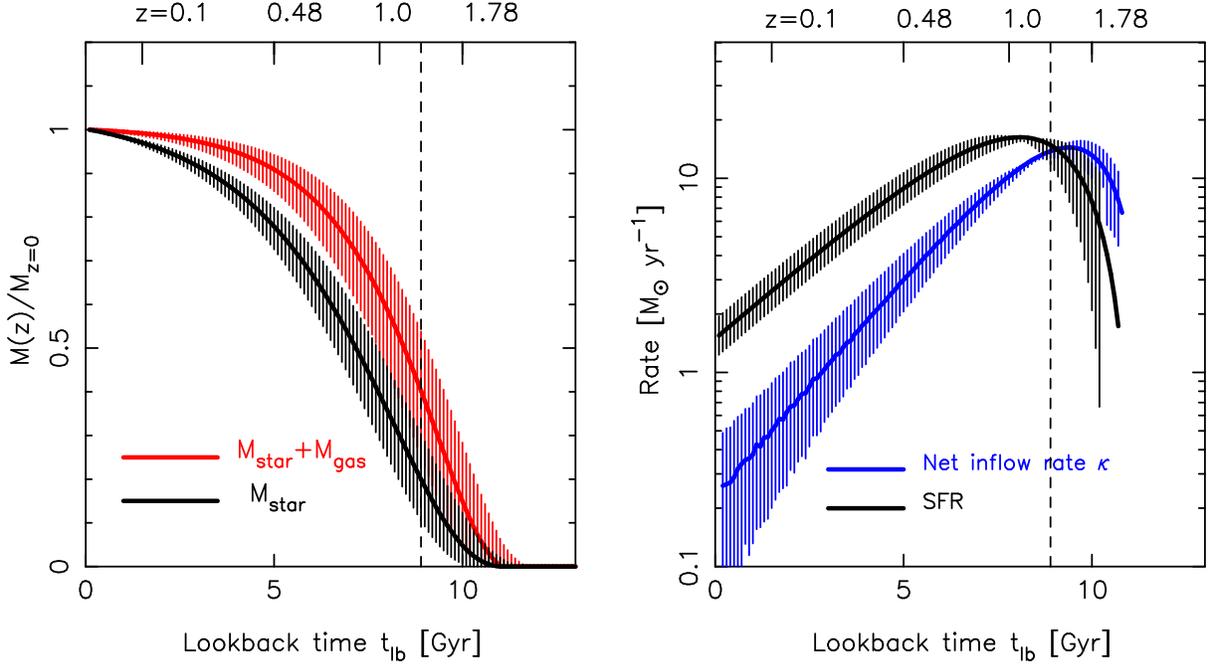}
\caption{Left: growth curves of $M_{\ast}$ (black line) and $M_{\ast}+M_{\rm gas}$ (red line). The uncertainties are calculated allowing variations of $\pm0.1$ dex in SFR, $\pm0.1$ dex in $M_{\rm H2}$ and $\pm0.2$ dex in $M_{\rm HI}$, respectively. The dashed line indicates $z=1.3$, beyond which the growth history of galaxies becomes very uncertain. With the $M_{\ast}+M_{\rm gas}$ growth curve, we can quantify net gas inflow rate $\kappa$ for progenitor galaxies, as shown in equation (7). Right: The evolution of the SFR (black line) and $\kappa$ (blue line).  }\label{fig3}
\end{figure*}
Direct determination of the \Hi~mass of galaxies ($M_{\rm HI}$) at $z>0.3$ is currently not realistic. Recently, \citet{Popping 2015} used an indirect technique to infer the evolution of the cold gas of SFGs from $z=3$ to $z=0.5$, finding that at fixed $M_{\ast}$, the $M_{\rm HI}$ of SFGs shows no redshift dependence. In the local Universe, deep \Hi~surveys such as GASS \citep{Catinella 2010, Catinella 2013} have compiled a representative galaxy sample to enable a direct investigation of the \Hi~mass for typical massive SFGs. In Figure \ref{fig2}, we compare the $M_{\rm HI}-M_{\ast}$ relations of \citet{Popping 2015} with that of the GASS sample. For the GASS galaxies, only those with both clear star formation (NUV$-r<4.0$) and \Hi~detection are selected. It can be seen that these two data sets show very good consistency. For the GASS galaxies, we fit the $M_{\rm HI}-M_{\ast}$ relation with:
\begin{equation}
{\rm log}(M_{\rm HI}/M_{\sun})=0.55{\rm log}(M_{\ast}/M_{\sun})+4.11,
\end{equation}
as shown in the black solid line in Figure~\ref{fig2}.

In what follows we assume that the $M_{\rm HI}-M_{\ast}$ relation has no evolution at $z=[0, 1.3]$, and $M_{\rm gas}$ is referred as $M_{\rm gas}=M_{\rm HI}+M_{\rm H2}$.

\section{The growth history of $M_{\ast}+M_{gas}$ and the inferred net gas inflow rate}
In the left panel of Figure \ref{fig3}, we show the growth curves of $M_{\ast}$ and $M_{\ast}+M_{\rm gas}$ for progenitor galaxies. It can be seen that $M_{\ast}+M_{\rm gas}$ grows much faster than $M_{\ast}$. At $z=0.5$, $M_{\ast}+M_{\rm gas}$ has assembled $\sim$90\% of its final mass, whereas only $\sim$75\% of the final stellar mass is assembled. Since $M_{\ast}$ contributes to the majority of the total baryonic budget at most epoches \textbf{($z<1$)}, the uncertainty of the $M_{\ast}+M_{\rm gas}$ growth curve is thus dominated by the uncertainty in the $M_{\ast}$ determination, i.e, the star formation history.

With the growth curve of $M_{\ast}+M_{\rm gas}$ in hand, we can quantify the evolution of net inflow rate $\kappa$. In a specific time interval of $\Delta t=t-t_{0}$, the net inflow mass is:
\begin{equation}
M_{\rm net}=M_{\rm inflow}-M_{\rm outflow},
\end{equation}
where $M_{\rm inflow}$ and $M_{\rm outflow}$ are the inflow and outflow mass during $\Delta t$, respectively. From mass conservation it is straightforward that
\begin{equation}
(M_{\ast}+M_{\rm gas})_{t}=(M_{\ast}+M_{\rm gas})_{t0}+M_{\rm net}.
\end{equation}
Then the net inflow rate $\kappa$ can be written as:
\begin{equation}
\kappa=\frac{M_{\rm net}}{\Delta t}=\frac{(M_{\ast}+M_{\rm gas})_{t}-(M_{\ast}+M_{\rm gas})_{t0}}{\Delta t}.
\end{equation}

In a more standard form, the brackets of equation (6) and (7) should include the mass of ionized gas and dust. However, in SFGs the mass of dust and ionized gas are both around two orders of magnitude lower than the mass of cold gas \citep{Wolfire 2003,Remy 2014}. Therefore, ignoring these two components should be safe. In the right panel of Figure \ref{fig3}, we show SFR and $\kappa$ as functions of lookback time $t_{\rm lb}$. Some interesting information can be read from this panel. First, the SFR reaches the peak value later than $\kappa$. This is comprehensible since to trigger star formation, the accreted gas needs to be further condensed. Second, the SFR declines by a factor $\sim \times10$ from $z=1.3$ to $z=0$, while at the same period $\kappa$ declines by a factor of $\sim \times50$. At $z\sim 0$, such a low net inflow rate ($\kappa \sim0.15\times$SFR) is far from sufficient to sustain the observed SFR. The fuel required for star formation in the present-day $M_{\rm MW}$ SFGs is thus mostly from internal sources, such as the recycled gas \citep{Leitner 2011} and the remaining gas reservoir.

We note that GASS is a very deep \Hi~survey, and the $M_{\rm HI}-M_{\ast}$ relation of GASS may be biased to gas-poor SFGs. To test how the $M_{\rm HI}-M_{\ast}$ relation impacts on our result, we have also applied the $M_{\rm HI}-M_{\ast}$ relation of the ALFALFA sample \citep{Giovanelli 2005} in our analysis. The ALFALFA survey is biased to \Hi~rich galaxies, as demonstrated in \citet{Huang 2012}. At log$(M_{\ast}/M_{\sun})>$10.0, the ALFALFA galaxies are systematically around $0.2$ dex more rich in \Hi~mass than the GASS galaxies. When applying the $M_{\rm HI}-M_{\ast}$ relation of ALFALFA, we found that the results are not changed. This is because $M_{\rm HI}$ only contributes to the minority of the baryonic mass budget (<30\%) at log$M_{\ast}/M_{\sun}>10.0$ even when the $M_{\rm HI}-M_{\ast}$ relation of ALFALFA is applied, thus having little impact on the $M_{\ast}+M_{\rm gas}$ growth curve. We thus conclude that a slight modification on the $M_{\rm HI}-M_{\ast}$ relation will not have a significant impact on our results.

\section{Constraining the inflow and outflow rates}
The gas phase metallicity, $Z_{\rm gas}$, can provide valuable insights in constraining the outflow properties of galaxies \citep{Finlator 2008, Lilly 2013,Belfiore 2016}.  In this section we will compare the observed $Z_{\rm gas}$ evolution of progenitor galaxies with that from a chemical evolution toy model to set constraints on the outflow rate $\zeta_{\rm outflow}$. Once $\zeta_{\rm outflow}$ is known, then we can investigate the inflow rate of these galaxies as the net inflow rate $\kappa$ has been determined.

For a galaxy that with a known $M_{\ast}$ growth history, its $Z_{\rm gas}$ at different redshifts can be inferred by utilizing the observed $M_{\ast}-Z_{\rm gas}$ relation (MZR)\citep{Maiolino 2008,Zahid 2013,Zahid 2014}. However, deriving the $Z_{\rm gas}$ evolution in this way may suffer large uncertainties, since different authors derive the MZRs using different sample section criteria and metallicity calibrations. To derive $Z_{\rm gas}$ in a consistent way across the probed redshift range, we infer $Z_{\rm gas}$ utilizing the tight correlation between $Z_{\rm gas}$, $M_{*}$ and SFR established at $z=0$. Based on the large $z=0$ SFG sample, \citet{Mannucci 2010} found that there exists a tight correlation among these three quantities, which can be expressed as:
\begin{equation}
12+{\rm log(O/H)}=8.90+0.39x-0.20x^2-0.077x^3+0.064x^4,
\end{equation}
where $x={\rm log}(M_{\ast})-0.32{\rm log(SFR)}-10$.

\begin{figure}
\centering
\includegraphics[width=80mm,angle=0]{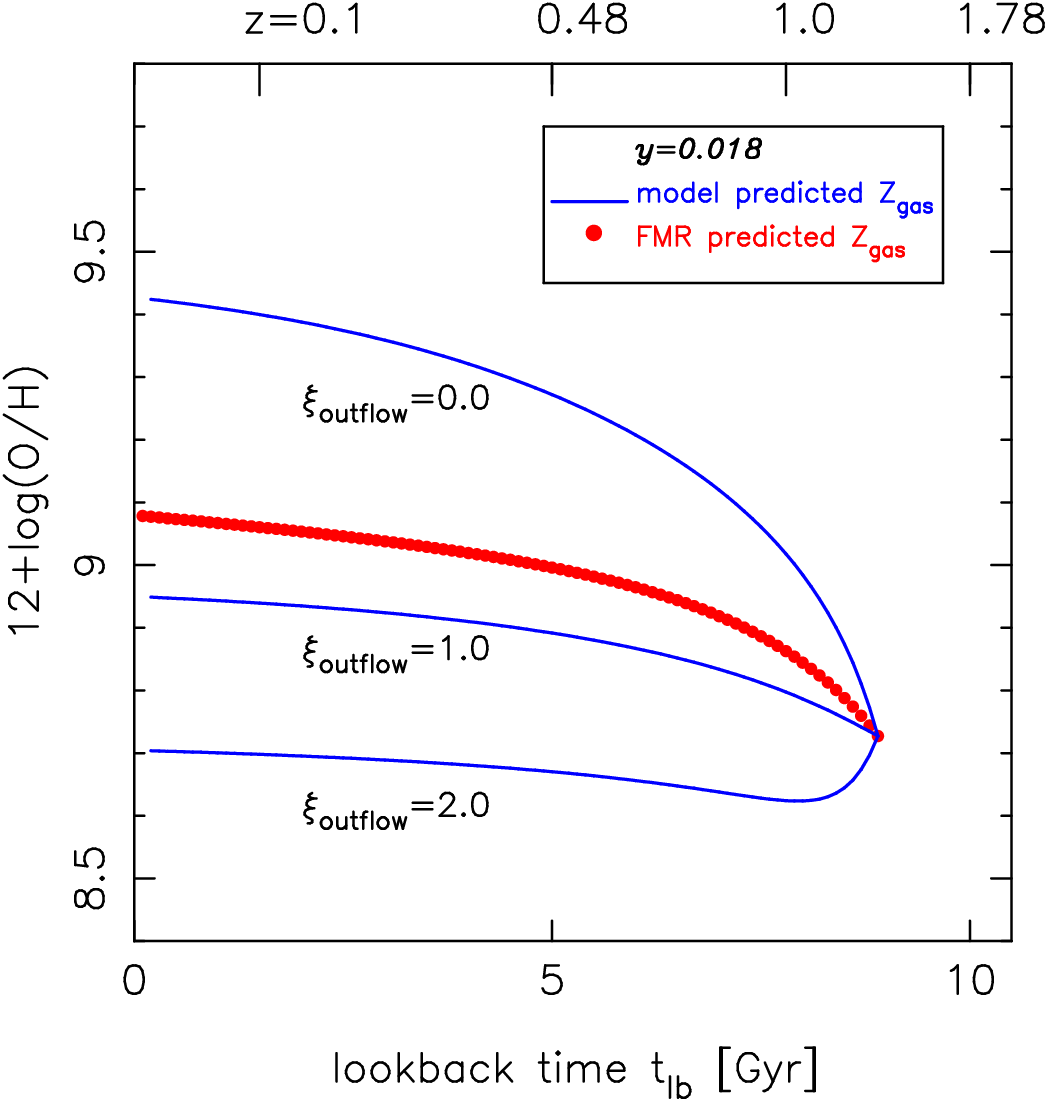}
\caption{For illustration, we compare the FMR predicted $Z_{\rm gas}$ evolution from $z=1.3$ to $z=0$ for progenitor galaxies (red symbols) with 3 examples from our analytical chemical evolution model (blue lines). In each model, the nucleosynthetic yield is fixed to $y=0.018$. }\label{fig4}
\end{figure}

\citet{Mannucci 2010} found that galaxies at $z<2.5$ appear all follow this relation, which they termed as the fundamental metallicity relation (FMR). There have been many recent studies investigating whether this $M_{\ast}-{\rm SFR}-Z_{\rm gas}$ relation evolves from high-$z$ to low-$z$. At $z<1.5$, the FMR seems do not evolve \citep{Cresci 2012,Yabe 2014}. At higher redshift, some studies report a same FMR as that established at $z=0$ \citep{Henry 2013,Maier 2014}, while some studies reported a possible redshift evolution in this relation \citep{Salim 2015,Sanders 2015,Sanders 2018}. Since this work focuses on the evolution of $M_{\rm MW}$ progenitor SFGs at $z<1.3$, we assume that the FMR does not evolve during this epoch. Inserting the $M_{\ast}$ and SFR of progenitor galaxies into equation (8), we derive the $Z_{\rm gas}$ evolution, as shown in the red symbols of Figure~\ref{fig4}.

By making some simple assumptions, the evolution of $Z_{\rm gas}$ can be derived analytically. By definition, $Z_{\rm gas}=M_{\rm Z, gas}/M_{\rm gas}$, where $M_{\rm Z, gas}$ is the mass of metals in the gas reservoir. For a given SFG, $M_{\rm Z, gas}$ can increase by the input of metals from star formation and metal-enriched inflows, or it can decrease by gas outflows and the lockup of metals into long-live stars. Assuming the inflow gas has a metallicity $Z_{0}$ and the metal produced by star formation is $y\times\rm SFR$ (where $y$ is the nucleosynthetic yield per stellar population), from the mass conservation of metals, the change of $M_{\rm Z, gas}$ per unit time, $dM_{\rm Z,gas}/dt$, can be written as:
\begin{equation}
\frac{dM_{\rm Z, gas}}{dt}=(y\cdot{\rm SFR}+Z_{0}\cdot\zeta_{\rm inflow})-Z_{\rm gas}\cdot\zeta_{\rm outflow}-Z_{\rm gas}(1-R)\cdot{\rm SFR},
\end{equation}
where $\zeta_{\rm inflow}$ and $\zeta_{\rm outflow}$ are gas inflow and outflow rate, and $R$ is the return mass fraction (defined as $R=mass~loss~rate/{\rm SFR}$), respectively. The last term represents the metal that locked in long-live stars. Following the definition of $Z_{\rm gas}$, then
\begin{equation}
\frac{dZ_{\rm gas}}{dt}=y\frac{\rm SFR}{M_{\rm gas}}-(Z_{\rm gas}-Z_{0})\frac{\zeta_{\rm inflow}}{M_{\rm gas}}.
\end{equation}
Assuming that $\zeta_{\rm inflow}$, $M_{\rm gas}$, $y$ and SFR are all constant or only change slowly during the time interval $\Delta t=t-t_{0}$, then the solution of equation (10) is:
\begin{equation}
Z_{\rm gas}(t)=Z_0+y\frac{{\rm SFR}}{\zeta_{\rm inflow}}+[Z_{\rm gas}(t_0)-Z_0-y\frac{{\rm SFR}}{\zeta_{\rm inflow}}]e^{{-\frac{\zeta_{\rm inflow}}{M_{\rm gas}}}(t-t_0)},
\end{equation}
as given by \cite{Peng 2014}.

When presenting the outflow rate $\zeta_{\rm inflow}$ in units of SFR:
\begin{equation}
\zeta_{\rm inflow}=\xi_{\rm outflow}\cdot{\rm SFR},
\end{equation}
$\zeta_{\rm inflow}$ can be written as:
\begin{equation}
\zeta_{\rm inflow}=\xi_{\rm outflow}\cdot{\rm SFR}+\kappa.
\end{equation}
Since the SFR, $M_{\rm gas}$ and $\kappa$ of progenitor galaxies have been derived in the above sections, given a starting $Z_{\rm gas}(t_0)$ and a set of input parameter $(Z_{0},~y,~\xi_{\rm outflow})$, one can predict the evolution of $Z_{\rm gas}$ at according to equation (11).

We first assume that the inflow gas is pristine, i.e., $Z_{0}=0$. This is a common assumption taken in most metallicity evolution models. The nucleosynthetic yield, $y$, is taken as a fixed value depending on the adopted IMF. In literatures, $y$ is around $0.01-0.05$ (see \citealt{Vincenzo 2016}, and references therein). The mass loading factor $\xi_{\rm outflow}$ is dependent on stellar mass. However, in the mass range considered, i.e., log$(M_{\ast}/M_{\sun})=10.0-10.7$, the dependence of $\xi_{\rm outflow}$ on mass is quite weak \citep{Peeples 2011}. We thus assume it to be a constant as well.

\begin{figure}
\centering
\includegraphics[width=90mm,angle=0]{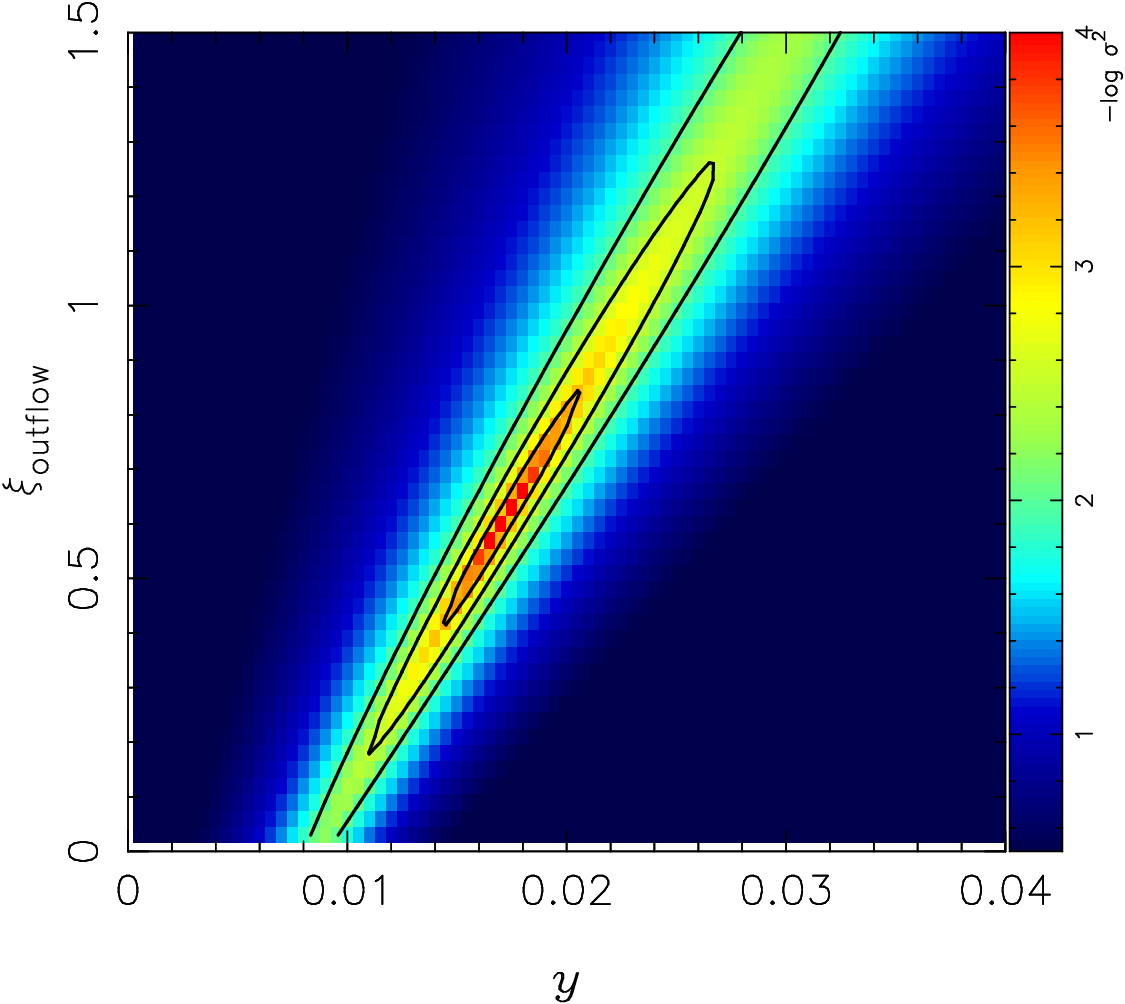}
\caption{The $\sigma^2$ map against $y$ and $\xi_{\rm outflow}$. It is clear that the lowest $\sigma^2$, i.e., the best matches between model predictions and observations, is found at $\xi_{\rm outflow}\sim 0.5-0.8$ and $y\sim 0.015-0.02$.}\label{fig5}
\end{figure}

With these simplifications, we predict the $Z_{\rm gas}$ evolution of progenitor galaxies at $z<1.3$, with a time interval of $\Delta t=0.1$ Gyr. At $z=1.3$ where the progenitor galaxy has log$(M_{\ast}/M_{\sun})=10.0$, $Z_{\rm gas}$ is log(O/H)+12=8.72 as predicted by the FMR. In Figure~\ref{fig4}, we show three examples of the $Z_{\rm gas}$ evolution curves predicted by our toy model, adopting $y=0.018$ and three different $\xi_{\rm outflow}$. A nucleosynthetic yield of $y=0.018$ is chosen because model predictions best match observations near this value, as shown below. As can be seen, the $Z_{\rm gas}$ evolution predicted by our model is quite sensitive to $\xi_{\rm outflow}$.

For a given parameter pair $(y,~\xi_{\rm outflow})$, we characterize the degree of the matching between model prediction and observation with:
\begin{equation}
\sigma^2=\sum\limits_{i=1}^{N}(\frac{Z_{\rm gas, model}(t_i)-Z_{\rm gas, FMR}(t_i)}{Z_{\rm gas, FMR(t_i)}})^2/N.
\end{equation}

\begin{figure}
\centering
\includegraphics[width=80mm,angle=0]{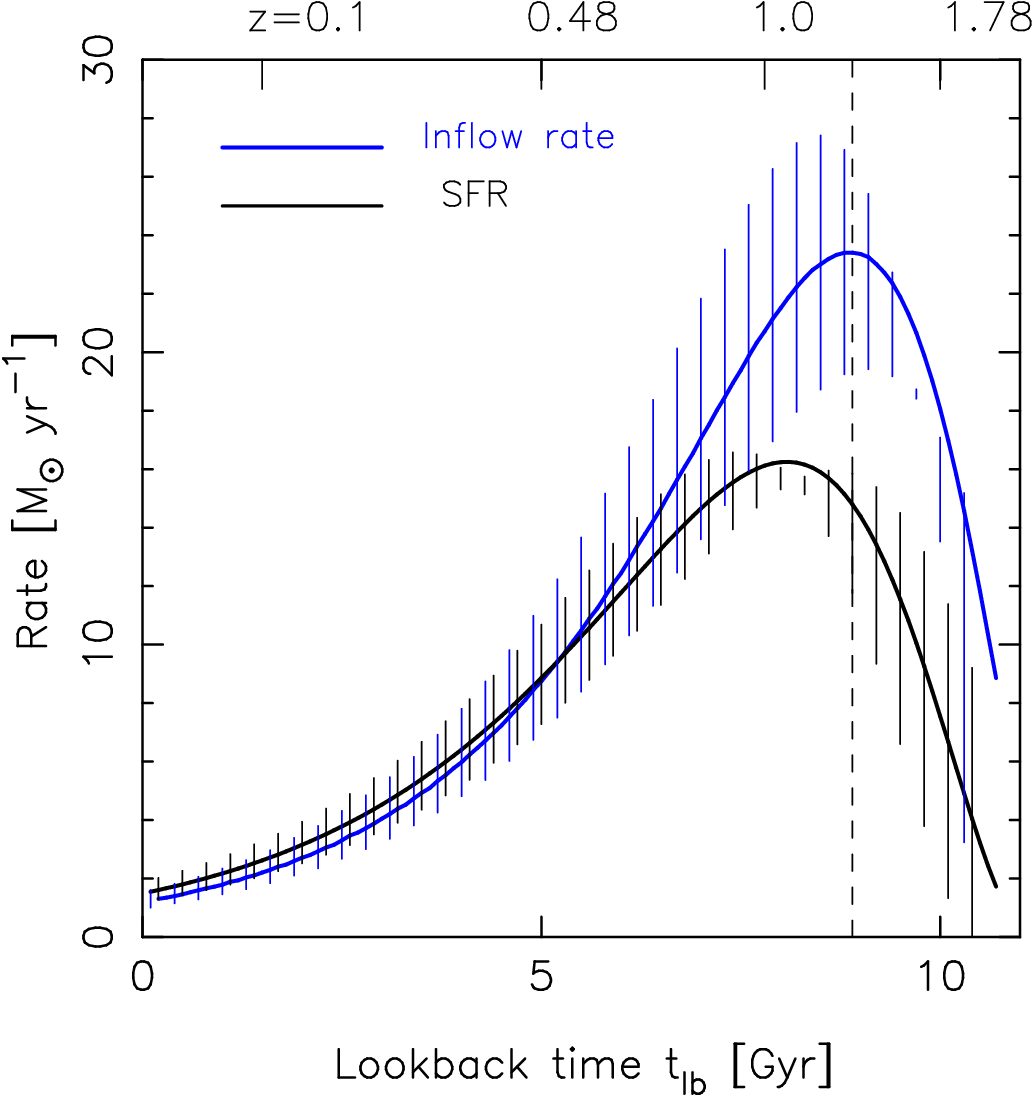}
\caption{The inferred gas inflow rate $\zeta_{\rm inflow}$ evolution (blue line), applying $\xi_{\rm outflow}=0.65$. Uncertainties are calculated by including those of the SFR and $\kappa$, and $\pm0.15$ in $\xi_{\rm outflow}$. We also plot the evolution of SFR for comparison. The dashed line indicates $z=1.3$.}\label{fig6}
\end{figure}

In Figure~\ref{fig5}, we show the $\sigma^2$ map against $y$ and $\xi_{\rm outflow}$. As can be seen, the best-fit mass loading factor is $\xi_{\rm outflow}\sim 0.5-0.8$. We insert the median value, $\xi_{\rm outflow}=0.65$, into equation (13) to derive the inflow rate $\zeta_{\rm inflow}$. In Figure~\ref{fig6}, we show the evolution of $\zeta_{\rm inflow}$ at $z<1.3$. At first glance, the evolution of $\zeta_{\rm inflow}$ can be largely divided into two phases: a rapidly declining phase at $0.5<z<1.3$, and a slowly evolving phase at $z<0.5$. For comparison we also plot the evolution of SFR in Figure~\ref{fig6}. At $z>1.0$, $\zeta_{\rm inflow}$ is clearly higher than the SFR. This may correspond to the ``gas accretion epoch" as predicted by theory \citep{Keres 2005}. At $z<0.5$, the SFR largely mimics the evolution of $\zeta_{\rm inflow}$, suggesting that the $M_{\rm MW}$ progenitor SFGs gradually enter a ``quasi-steady" phase at late epoches during which their SFRs are self-regulated by the balance between gas inflows and outflows.

\section{Discussion}
Combining the observed evolution of the SFR$-M_{\ast}$ and $M_{\rm gas}-M_{\ast}$ relation of SFGs, we constrain the gas flow histories of local Milky Way-mass SFGs since $z=1.3$. Below we will compare our results with previous works and discuss the implications of these results.

At $z\sim 1.3$, we find that the net inflow rate $\kappa$ reaches a level comparable to the SFR (right panel of Figure~\ref{fig3}). \citet{Papovich 2011} also reported a similar phenomenon for galaxies at constant number density of $n=2\times10^{-4}~ \rm Mpc^{-3}$ at $z\sim 3.0$ (see their Figure 4). Recently, \citet{scoville 2017} investigated the evolution of $\kappa$ for SFGs since $z\sim 3.0$. The authors found that the ratio between $\kappa$ and SFR, $\kappa$/SFR,  closely correlates with $z$ and $M_{\ast}$:
\begin{equation}
\kappa/{\rm SFR}\sim (1+z)^{0.7}(M_{10}^{0.56}-0.56 \times M_{10}^{0.74}),
\end{equation}
where $M_{10}$ is the stellar mass in units of $10^{10}M_{\sun}$. According to equation (15), SFGs with log$(M_{\ast}/M_{\sun})=10.0$ typically have $\kappa$/SFR$\sim$0.85 at $z\sim 1.3$, which is in good agreement with ours.

At $z\sim 0$, we find a very low net inflow rate, $\kappa \sim 0.15\times$SFR, for the Milky Way-mass SFGs. This is lower than that reported in \citet{scoville 2017}, $\kappa \sim 0.4\times$SFR. We emphasize that this discrepancy is largely due to the different treatments on mass loss rate applied in these two works. From equation (11) of \citet{scoville 2017}, it is clear that the derived $\kappa$ is directly coupled with the applied return mass fraction $R$, in the sense that a large $R$ will yield a small $\kappa$. In \citet{scoville 2017}, the authors used a constant return mass fraction of $R=0.3$ across $z=0-3$. In this work, we use the full Main Sequence Integration approach, in which $R$ is not a constant but will increase towards low redshifts, because mass loss contributed from old stellar populations becomes increasingly important at late epochs. At $z\sim 0$, the MSI-based return mass fraction is $R\sim0.6$, which in turn results in a reduction of $\sim 0.3\times$SFR in $\kappa$ compared to that of \citet{scoville 2017}.

We find the best-fit mass loading factor is $\xi_{\rm outflow}\sim 0.5-0.8$. Although this is quantitatively consistent with that found in some previous works \citep{Lilly 2013,Yabe 2015,Belfiore 2016}, it is worthy to note that $\xi_{\rm outflow}$ and $y$ are degenerated in our chemical evolution model, as shown in Figure~\ref{fig5} (also see \citealt{Peeples 2011}). As such, the derived $\xi_{\rm outflow}$ is highly sensitive to the choice of $y$: one must know how many metals are produced before he/she can determine the level of outflows required to produce the evolution of $Z_{\rm gas}$. To set more stringent constraints on $\xi_{\rm outflow}$, complementary approaches are thus needed.

There appears to be a ``turnover" in the gas inflow rate evolution curve at $z\sim 0.5$ (Figure~\ref{fig6}). Specifically, at $z=0.5-1.3$, the change rate of $\zeta_{\rm inflow}$, $\frac{d \zeta_{\rm inflow}}{dt}$, is relatively stable with $\frac{d \zeta_{\rm inflow}}{dt} \sim4.0~M_{\sun}~Gyr^{-2}$, whereas at $z<0.5$ this rate is only $\sim$ $1.3~M_{\sun}~Gyr^{-2}$. What is the physics behind this phenomenon? Under the current framework of galaxy formation, we speculate that this turnover may reveal a switch from the ``cold mode" to the ``hot mode" accretion near the critical halo mass $M_{\rm c}$. Interestingly, at the ``turnover" redshift the progenitor galaxies have log$(M_{\ast}/M_{\sun})\sim 10.6$, corresponding to a halo mass of $M_{\rm h}\sim 1\times 10^{12}M_{\sun}$ \citep{Behroozi 2013}. This is well consistent with the prediction of the halo-shock heating scenario.

To investigate whether other galaxies also exhibit a similar turnover in $\zeta_{\rm inflow}$ at a same $M_{\rm h}$, we also study SFGs of two different stellar masses (see Figure~\ref{fig9} of the Appendix). For an SFG that with a final stellar mass of log$(M_{\ast}/M_{\sun})= 11.0$, we find a similar turnover in its $\zeta_{\rm inflow}$ at $z\sim0.7$, at which the stellar mass of the progenitor galaxy is around log$(M_{\ast}/M_{\sun})=10.8$. Since these two turnover redshifts are only slightly different, we apply a same $M_{\rm h}-M_{\ast}$ relation to this galaxy, finding that the corresponding turnover halo mass is $M_{\rm h}\sim 2\times10^{12}M_{\sun}$. We argue that the turnover in $\zeta_{\rm inflow}$ doesn't occur at a same $M_{\rm h}$ for different SFGs. Interestingly, we note that the evolution trend of $M_{\rm turnover}$ is similar to that of $M_{\rm transition}$ \footnote{$M_{\rm transition}$ is the stellar mass at which the fraction of quenched galaxies reaches $f_{\rm quenched}=50\%.$} as reported in \citet{Haines 2017} (see their Figure 4). This may suggest a connection between the cessation of star formation in galaxies and the significant change in their gas inflow behaviors, as we will argue below.

Although this work is focused on the gas flow behavior of SFGs, our results may provide some insights in interpreting the star formation quenching of galaxies near or above $M^{\ast}$ (``mass quenching", see \citealt{Peng 2010}). Since the tight SFR$-M_{\ast}$ relation exists up to at least $z\sim 5-6$ \citep{speagle 2014,Tasca 2015}, the progenitors of massive quenched galaxies are expected to be normal SFGs before they get quenched. As such, a quenched galaxy should also experience a ``rapidly declining phase" in $\zeta_{\rm inflow}$ during a certain epoch. When $\zeta_{\rm inflow}$ has significantly decreased, the impact of internal processes on galaxy evolution will become increasingly important. It has been suggested that violent bulge build-up processes are often accompanied with gas outflow driven by strong starburst or AGN activities (or both), which is expected to be capable in cleaning the gas reservoir in a relatively short time scale \citep{Hopkins 2006, Geach 2014, Geach 2018}. Since $\zeta_{\rm inflow}$ has significantly decreased and the gas replenishment time scale is long, the removal of gas reservoir may drive the galaxy rapidly get quenched. Observationally, rapidly quenching systems (known as ``post-starburst" galaxies) are found to be bulge-dominated and with a surprisingly high AGN fraction, supporting this scenario \citep{Vergani 2010,Yesuf 2014,Baron 2018}. On the other hand, bar-driven bulge build-up processes may also play an important role in exhausting the cold gas reservoirs, although the timescale is relatively long \citep{Masters 2012,Wang 2012, Cheung 2013, Gavazzi 2015,Lin 2017}.

When a prominent bulge has been formed, other internal processes may also play a role in further suppressing star formation. Using cosmological simulations, \citet{Martig 2009} have illustrated that a prominent bulge is able to stabilize the gas disk against fragmentation to form stars. Recently, such kinds of dynamically driven star formation suppression are reported in observational studies \citep{Davis 2014,Genzel 2014b}. In addition, a bulge will play a role in preventing the the cooling of recycled gas. This is because in dispersion-supported (spheroidal) systems, a considerable fraction of the recycled gas will quickly mix with halo gas \citep{Parriott 2008}. By contrast, in disk-dominated galaxies the recycled gas can directly return to the co-rotating interstellar medium to form next-generation stars. Finally, winds driven by low-level AGNs appear capable in heating the surrounding gas to prevent star formation at the late epoches of galaxy evolution \citep{Cheung 2016,Weinberger 2017,Weinberger 2018,Li 2018}. In summary, we suggest that a significant decline in gas inflow rate is the first step required to quench a massive galaxy. Once this happens, bulge-related internal processes likely play an important role in quenching star formation, resulting in the strong correlation between sSFR and surface mass density \citep{Bell 2008, Franx 2008,Bell 2012,Cheung 2012,Fang 2013, Barro 2017,Whitaker 2017}.

\section{Summary and conclusions}
In this paper, we study the gas flow histories for the progenitors of local Milky Way-mass star-forming galaxies out to $z\sim1.3$. Assuming that the progenitor galaxies grow in their stellar mass mainly via star formation (not via mergers), then their stellar mass growth histories can be traced following the evolution of the SFR$-M_{\ast}$ relation. Using the molecular gas scaling relations established by \citet{Tacconi 2018}, we derive the molecular gas mass of progenitor galaxies. The \Hi~ gas mass is estimated based on the $M_{\rm HI}-M_{\ast}$ relation established at $z=0$, assuming that this relation does not evolve out to $z=1.3$. With the $M_{\ast}+M_{\rm gas}$ growth curve and chemical evolution modeling of progenitor galaxies, we have found the following:

1. From $z=1.3$ to $z=0$, the net inflow rate $\kappa$ decreases by a factor of $\sim \times$50, whereas the SFR decreases $\sim \times10$. At $z=0$, $\kappa$ is only $\sim0.15\times$SFR.

2. The mean outflow rate is $\sim (0.5-0.8)\times {\rm SFR}$.

3. The inflow rate $\zeta_{\rm inflow}$ experiences a ``rapidly declining phase" at $z=0.5-1.3$, during which $\zeta_{\rm inflow}$ decreases by $\sim 80$\%. At $z<0.5$, $\zeta_{\rm inflow}$ continuously decreases but with a much lower decreasing rate.

We suggest that when the gas inflow rate has significantly decreased, bulge-related internal processes likely play an important role in quenching star formation.

\acknowledgments
We thank the anonymous referee for constructive suggestions that help improving the clarity of the manuscript. This work was partially supported by the National Key Research and Development Program (``973" program) of China (No.2015CB857004, 2016YFA0400702, 2017YFA0402600 and 2017YFA0402703), the National Natural Science Foundation of China (NSFC, Nos. 11773001, 11721303, 11703092, 11320101002, 11421303, 11433005, 11773076 and 11721303), and the Natural Science Foundation of Jiangsu Province (No.BK20161097).

\appendix\label{sec:app}

For comparison, we have also studied the gas inflow and outflow histories of SFGs that with final stellar masses lower or higher than $M_{\rm MW}$. Here we present the results of two SFGs, of which one with a final stellar mass of log$(M_{\ast}/M_{\sun})=10.3$ and the other with log$(M_{\ast}/M_{\sun})=11.0$. For the low-mass SFG, the application of our methodology should be safe. For the high-mass one, we assume that the growth of its stellar mass is also dominated by in-situ star formation (not by mergers) and our method is still valid. This is supported by the study of \cite{Moster 2013}, who found that mergers only contribute $<20$\% to the total stellar mass budget of a log$(M_{\rm h}/M_{\sun})=13.0$ halo.

\begin{figure}
\centering
\includegraphics[width=80mm,angle=0]{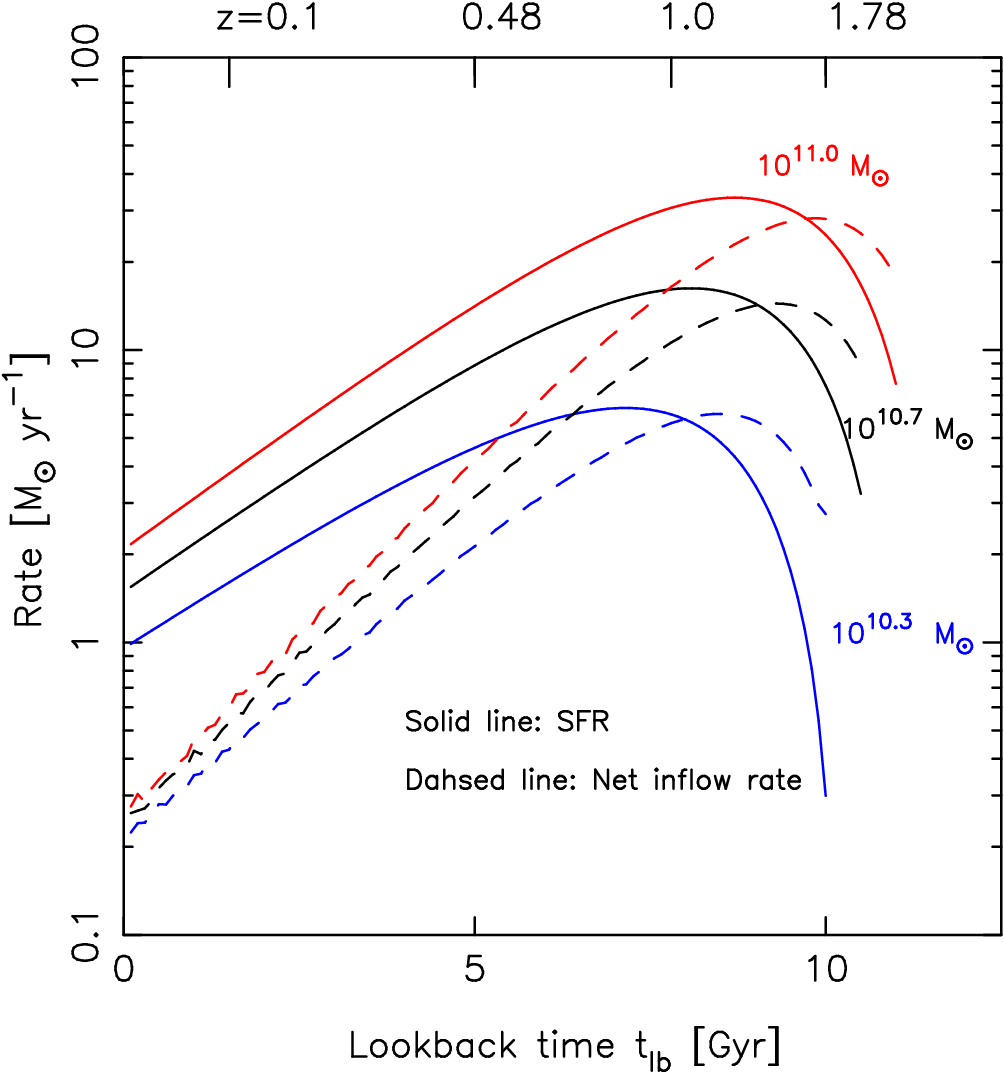}
\caption{SFR and net inflow rate as a function of cosmic time for galaxies with different masses. SFRs and net inflow rates are indicated in solid lines and dashed lines, respectively. Galaxies of different masses are indicated in different colors.}\label{fig7}
\end{figure}

The stellar mass growth histories and gas masses of these two galaxies are derived using the method described in Section 2 and Section 3. In Figure~\ref{fig7}, we compare the star formation as well as the net gas inflow histories of these two SFGs with those of the $M_{\rm MW}$ SFG. As can be seen, the star formation of high-mass SFGs peaks at higher redshifts than that of the low-mass ones. Another interesting feature is that the decreasing rate of net inflow rate is also mass dependent, with the most massive SFG has the highest net inflow decreasing rate.

\begin{figure}
\centering
\includegraphics[width=80mm,angle=0]{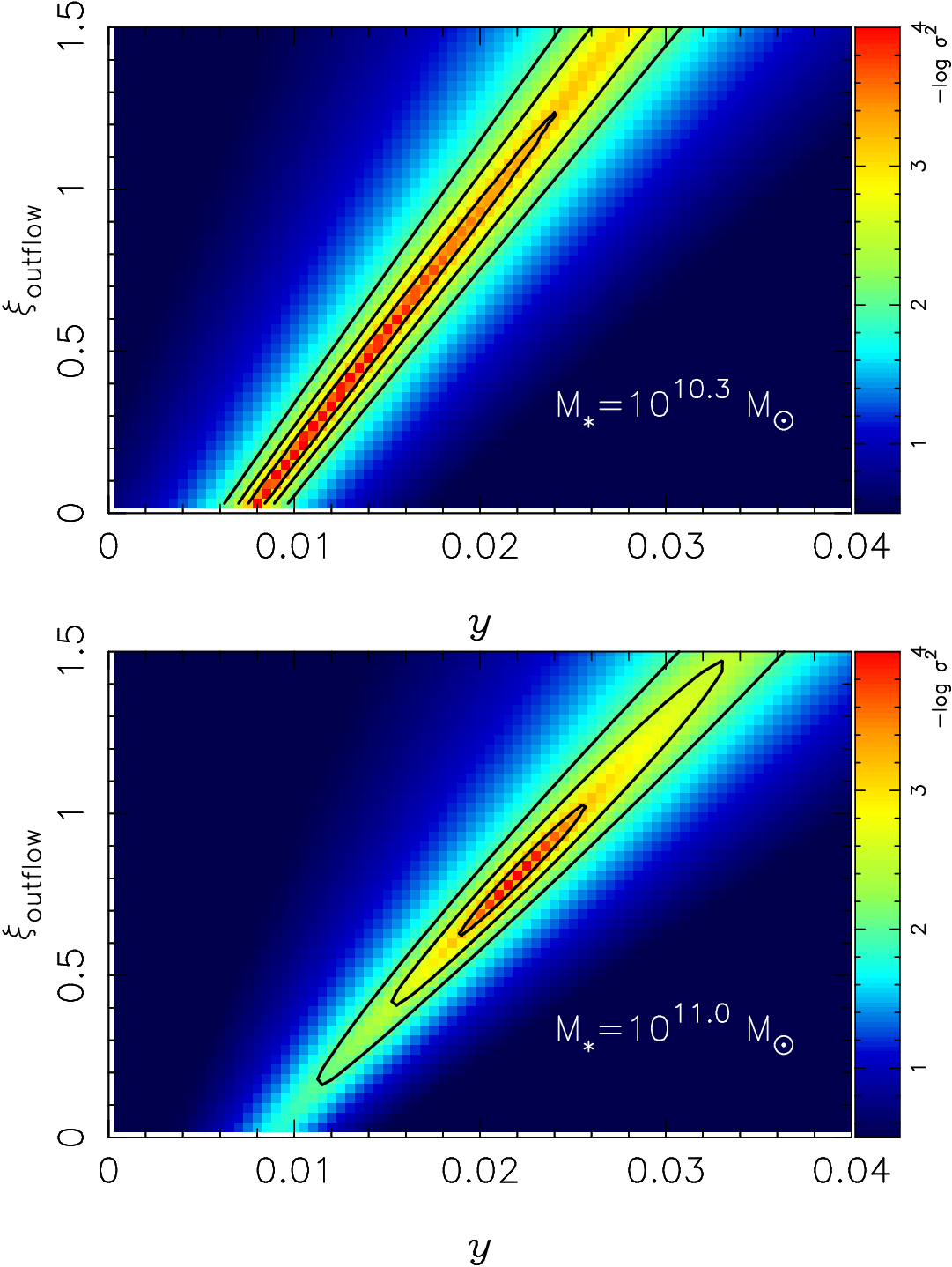}
\caption{Similar to Figure~\ref{fig5} but for SFGs with log$(M_{\ast}/M_{\sun})=10.3$ and log$(M_{\ast}/M_{\sun})=11.0$. }\label{fig8}
\end{figure}

We then model the $Z_{\rm gas}$ evolution of these two galaxies with the same method described in Section 5. The beginning redshifts are selected at which the galaxy has log$(M_{\ast}/M_{\sun})=10.0$. These correspond to $z=0.7$ and $z=1.9$ for the low-mass and high-mass SFG, respectively. In Figure~\ref{fig8}, we present the $\sigma$ map against $y$ and $\xi_{\rm outflow}$. As shown in the upper panel, $y$ and $\xi_{\rm outflow}$ can not be very well constrained for the low-mass galaxy, which is mainly due to the narrow redshift range available for the fitting procedure. For the high-mass galaxy, $y$ and $\xi_{\rm outflow}$ are better constrained, with the best fit $y\sim 0.023$ and $\xi_{\rm outflow}\sim 0.8$. Note that both $y$ and $\xi_{\rm outflow}$ are slightly higher than those derived for the $M_{\rm MW}$ SFG.

A comparison between Figure~\ref{fig5} and Figure~\ref{fig8} indicates that $\xi_{\rm outflow}$ may be mass dependent. This conflicts with our model assumption, that $\xi_{\rm outflow}$ is largely independent on stellar mass at log$(M_{\ast}/M_{\sun})>10.0$. We consider that this confliction may arise from the following aspects. First, the mass independence of $\xi_{\rm outflow}$ at log$(M_{\ast}/M_{\sun})>10.0$ is derived from the modeling of the mass-metallicity relation of low redshift SFGs \citep{Spitoni 2010,Peeples 2011}. For a certain SFG, it is difficult to determine whether $\xi_{\rm outflow}$ is roughly a constant during its evolution at log$(M_{\ast}/M_{\sun})>10.0$, because the $\xi_{\rm outflow}-M_{\ast}$ relation may have evolved from high$-z$ to low$-z$.  Second, the uncertainties of all input parameters, such as $M_{\rm gas}$, $\kappa$ and SFR, will more or less contribute to the output of $\xi_{\rm outflow}$. Finally, the scatter of the FMR, which is at a level of $\Delta$log$({\rm O/H})\sim 0.05$ dex \citep{Mannucci 2010}, is not taken into account during the fitting procedure. The combination of these factors may result in an offset between the output $\xi_{\rm outflow}$ and the true value. It is thus important to access whether the output $\xi_{\rm outflow}$ is reliable. When a same $y=0.018$ is adopted, the best-fit mass loading factor is $\xi_{\rm outflow}=0.8$ and 0.5 for the low-mass and high-mass SFG, respectively. To our knowledge, the difference between these two values is not significant, and we think that our initial model assumption (i.e., $\xi_{\rm outflow}=constant$) is still valid. Fixing $y=0.018$, we have also tried to use a mass-dependent $\xi_{\rm outflow}$, which is parameterized by $\xi_{\rm outflow}=a+b\times[{\rm log}(M_{\ast}/M_{\sun})-10.0]$, to set constraint on $\xi_{\rm outflow}$. We investigated the $\sigma$ map against $a$ and $b$ and found that the best-match is always near $b\sim -0.1$, i.e., $\xi_{\rm outflow}$ is indeed very weakly dependent on $M_{\ast}$ at log$(M_{\ast}/M_{\sun})>10.0$. In the following, we still assume that $\xi_{\rm outflow}$ is a constant across the mass range of log$(M_{\ast}/M_{\sun})=[10.0,11.0]$ and adopt a median value of $\xi_{\rm outflow}$=0.65.

\begin{figure}
\centering
\includegraphics[width=80mm,angle=0]{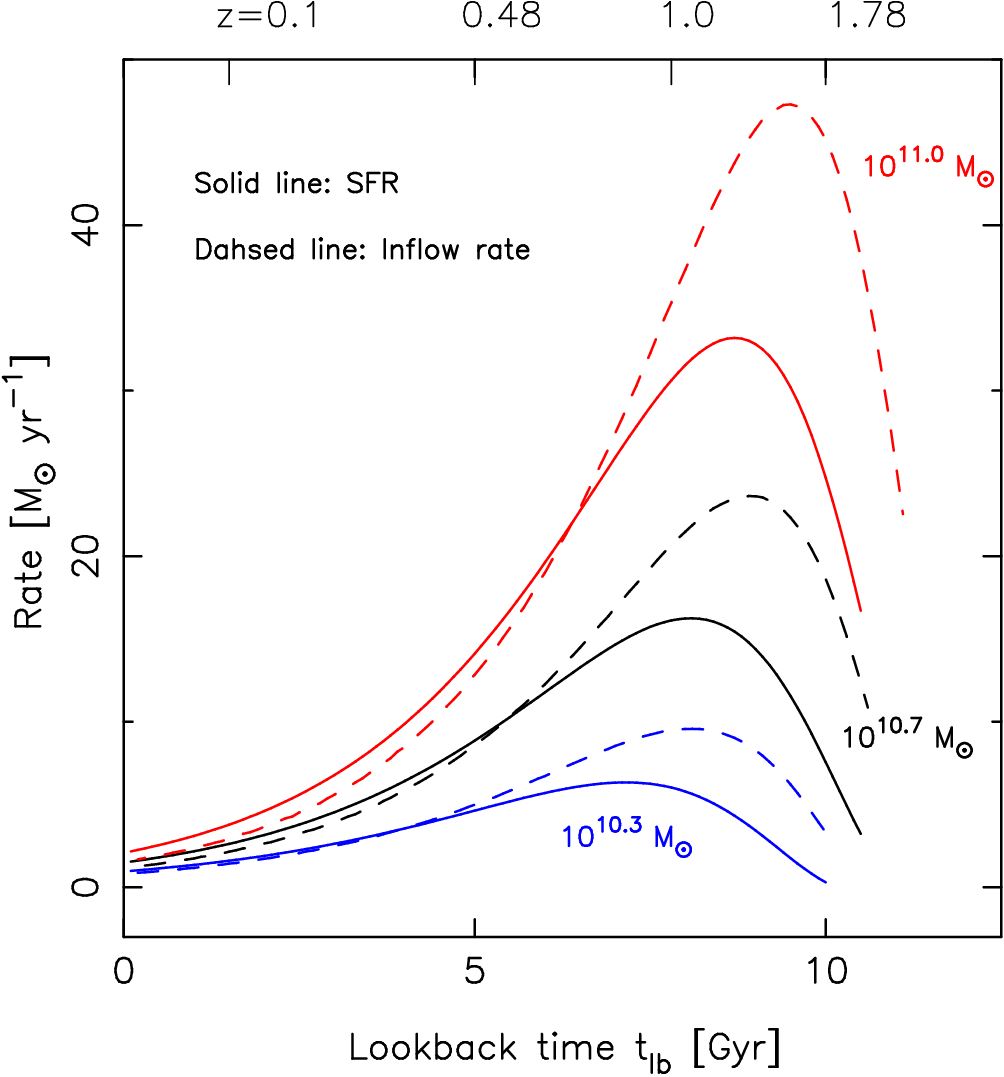}
\caption{Similar to Figure~\ref{fig6} but for SFGs with three different masses. }\label{fig9}
\end{figure}

In Figure~\ref{fig9}, we show the evolution of SFR and inflow rate for the three galaxies shown in Figure~\ref{fig7}, adopting $\xi_{\rm outflow}$=0.65. As can be seen, the inflow rates of these galaxies all exceed the SFRs in their early assembly epochs, and the inflow rates reach the peak values earlier than the SFRs by $\sim 1$ Gyr. Similar to Figure~\ref{fig6}, we define the epoch at which SFR$=\zeta_{\rm inflow}$ as the ``turnover" redshift.  As can be seen, the turnover redshift is $\sim0.7$ for the most massive SFG, at which the progenitor galaxy has log$(M_{\ast}/M_{\sun})\sim10.8$. At $z<z_{\rm turnover}$, the behaviors of $\zeta_{\rm inflow}$ are very close to the SFRs for all three SFGs.

\end{document}